\documentclass[floatfix,twocolumn,aps,pra,reprint,superscriptaddress]{revtex4-1}
\usepackage{graphicx}
\usepackage{amsmath}
\usepackage{amssymb}
\usepackage{hyperref}

\newcommand{\abs}[1]{\ensuremath{\left| #1 \right|}}

\newcommand{\grad}{\boldsymbol{\nabla}}
\newcommand{\pder}[2]{\frac{\partial#1}{\partial#2}}
\newcommand{\threevec}[3]{\left(\begin{array}{c}#1\\#2\\#3\end{array}\right)}

\newcommand{\threemat}[9]{\left(\begin{array}{ccc}#1&#2&#3\\#4&#5&#6\\#7&#8&#9\end{array}\right)}

\newcommand{\F}{\mathbf{F}}
\newcommand{\B}{\mathbf{B}}
\newcommand{\m}{\mathbf{m}}
\newcommand{\absF}{|\langle\mathbf{\hat{F}}\rangle|}
\newcommand{\expF}{\langle\mathbf{\hat{F}}\rangle}
\newcommand{\muB}{\ensuremath{\mu_\mathrm{B}}}
\newcommand{\SO}{\mathrm{SO}}

\newcommand{\rr}{\ensuremath{\mathbf{r}}}
\newcommand{\rrp}{\ensuremath{\mathbf{r}^\prime}}
\newcommand{\kk}{\ensuremath{\mathbf{k}}}

\newcommand{\rhat}{\ensuremath{\mathbf{\hat{r}}}}

\newcommand{\zhat}{\ensuremath{\mathbf{\hat{z}}}}
\newcommand{\nhat}{\ensuremath{\mathbf{\hat{n}}}}
\newcommand{\khat}{\ensuremath{\mathbf{\hat{k}}}}

\newcommand{\cdd}{\ensuremath{c_\mathrm{d}}}
\newcommand{\xidd}{\ensuremath{\xi_\mathrm{d}}}
\newcommand{\bb}{\ensuremath{\boldsymbol{\mathcal{B}}}}

\newcommand{\bcomp}{\ensuremath{\mathcal{B}}}
\newcommand{\Qtens}{\ensuremath{\mathsf{Q}}}
\newcommand{\QtensF}{\ensuremath{\mathsf{\tilde{Q}}}}
\newcommand{\Btens}{\ensuremath{\mathsf{B}}}
\newcommand{\BtensF}{\ensuremath{\mathsf{\tilde{B}}}}

\begin{document}

\author{Magnus O.\ Borgh}
\affiliation{Faculty of Science, University of East Anglia, Norwich, NR4 7TJ,
  United Kingdom}
\affiliation{Mathematical Sciences, University of Southampton,
  Southampton, SO17 1BJ, United Kingdom}
\author{Justin Lovegrove}
\affiliation{Mathematical Sciences, University of Southampton,
  Southampton, SO17 1BJ, United Kingdom}
\author{Janne Ruostekoski}
\affiliation{Mathematical Sciences, University of Southampton,
  Southampton, SO17 1BJ, United Kingdom}

\date{\today}

\title{Internal structure and stability of vortices in a dipolar spinor Bose-Einstein condensate}

\begin{abstract}
We demonstrate how dipolar interactions can have pronounced
effects on the structure of vortices in atomic spinor Bose-Einstein condensates
and illustrate generic physical principles that apply across dipolar spinor
systems. We then find and analyze the cores of singular vortices with
non-Abelian charges in the point-group symmetry of a spin-3 $^{52}$Cr
condensate. Using a simpler model system, we analyze the underlying dipolar
physics and show how a characteristic length scale arising from the magnetic
dipolar coupling interacts with the hierarchy of healing lengths of the $s$-wave
scattering, and leads to simple criteria for the core structure: When the
interactions both energetically favor the ground-state spin condition, such as
in the spin-1 ferromagnetic phase, the size of singular vortices is restricted
to the shorter spin-dependent healing length. Conversely, when the interactions
compete (e.g., in the spin-1 polar phase), we find that the core of a singular
vortex is enlarged by increasing dipolar coupling. We further demonstrate how
the spin-alignment arising from the interaction anisotropy is manifest in the
appearance of a ground-state spin-vortex line that is oriented perpendicularly
to the condensate axis of rotation, as well as in potentially observable
internal core spin textures.  We also explain how it leads to
interaction-dependent angular momentum in nonsingular vortices as a result of
competition with rotation-induced spin ordering. When the anisotropy is modified
by a strong magnetic field, we show how it gives rise to a symmetry-breaking
deformation of a vortex core into a spin-domain wall.
\end{abstract}

\maketitle

\section{Introduction}
\label{sec:introduction}

The achievement of Bose-Einstein condensation using atoms with large
magnetic dipole moments, such as
$^{52}$Cr~\cite{griesmaier_prl_2005,pasquiou_prl_2011,de-paz_pra_2013},
$^{168}$Er~\cite{aikawa_prl_2012},
and several Dy isotopes~\cite{lu_prl_2011,tang_njp_2015} as well as
creation of a degenerate dipolar Fermi gas~\cite{aikawa_prl_2014} have
opened up a
new avenue for studying the effects of long-range and
anisotropic interactions in ultracold atomic
gases~\cite{lahaye_rpp_2009}.  In such systems,
long-range magnetic order can coexist with superfluidity, making
possible, e.g., ferro-superfluids~\cite{lahaye_nature_2007}. The
interaction can then lead to novel instabilities, e.g., toward formation
of droplet
crystals~\cite{kadau_nature_2016,bisset_pra_2015,xi_pra_2016}, and
formation of a condensate may be strongly influenced by the spin
dynamics~\cite{naylor_prl_2016}.
The interaction can also profoundly affect the stability and structure of
vortices~\cite{yi_pra_2006b,o'dell_pra_2007,wilson_prl_2008,klawunn_prl_2008,abad_pra_2009},
e.g., inducing a phase transition from straight to twisted vortex
lines~\cite{klawunn_njp_2009}.
Simultaneously, the structure of
topological defects and textures is a central topic in the study of
spinor Bose-Einstein condensates (BECs), where the atomic spin degree
of freedom is not frozen out
by strong magnetic fields~\cite{kawaguchi_physrep_2012}. This gives rise
to a rich phenomenology of the internal structure of vortices~\cite{yip_prl_1999,mizushima_pra_2002,mizushima_prl_2002,martikainen_pra_2002,saito_prl_2006,ji_prl_2008,lovegrove_pra_2012,kobayashi_pra_2012,borgh_prl_2012,lovegrove_prl_2014,lovegrove_pra_2016,borgh_prl_2016}.
Recent
experiments have demonstrated controlled preparation of nonsingular
vortices~\cite{leanhardt_prl_2003,leslie_prl_2009,choi_prl_2012,choi_njp_2012}
as well as point defects~\cite{ray_nature_2014,ray_science_2015} and
particle-like solitons~\cite{hall_nphys_2016}.
The \emph{in situ} observation of splitting of singly quantized
vortices into pairs of half-quantum vortices~\cite{seo_prl_2015},
theoretically predicted in Ref.~\cite{lovegrove_pra_2012},
marks an increasing experimental interest in the internal core structure.

The spin degree of freedom in spinor BECs also implies that dipolar
interactions (DIs) arising
from the magnetic dipole moment of the atoms can have a strong impact
on the spin
texture~\cite{kawaguchi_prl_2007,kawaguchi_pra_2010,huhtamaki_pra_2010a}.
Nontrivial textures arising spontaneously due to DI have been observed
in experiment~\cite{vengalattore_prl_2008,eto_prl_2014}.  However, the
potentially large impact on the internal structure of vortices
has so far been little
studied.  Even a very weak DI can influence the relaxation of vortices by
changing the longitudinal magnetization, whose conservation
in $s$-wave scattering can be important for stability and
structure, e.g., of a coreless vortex in a polar
condensate~\cite{lovegrove_prl_2014,lovegrove_pra_2016}.
Theoretical works on vortices in dipolar spinor BECs have predicted a
superfluid
Einstein--de~Haas effect, where magnetic relaxation induces vortex
formation~\cite{santos_prl_2006,kawaguchi_prl_2006a,gawryluk_prl_2011,swislocki_pra_2011},
as well as a stable spin vortex in a nonrotating
system~\cite{yi_prl_2006,kawaguchi_prl_2006b,takahashi_prl_2007}. In a
rotating highly oblate condensate, complex multivortex states and
stable higher-order defects have been
described~\cite{simula_jpsj_2011}.

Here we demonstrate how dipolar interactions can have
pronounced effects on the internal structure of vortices in atomic
spinor BECs.
To clearly illustrate the underlying physical principles, we employ
the comparative simplicity of a spin-1 model system.  The dipolar effects
arise from generic properties of the DI and spinor systems and
the corresponding principles
may therefore be applied more broadly to understand properties of
vortices in dipolar spinor BECs.
Dipolar spin-1 BECs could potentially also be realized using
alkali-metal atoms  by suppressing the $s$-wave scattering lengths via
optical or microwave Feshbach
resonances~\cite{fatemi_prl_2000,papoular_pra_2010}. As an example of
experimentally realized dipolar BEC, we numerically find
and analyze the stable core structure of a singular
vortex in a spin-3 condensate of $^{52}$Cr. The $^{52}$Cr atom
possesses a relatively large magnetic dipole
moment~\cite{santos_prl_2006} and is
predicted to exhibit a dihedral-6 point-group order-parameter
symmetry in the ground state, supporting non-Abelian vortices.

The multicomponent condensate wave function of a spinor BEC allows the
condensate to maintain the superfluid density in the core of singular
vortices.  In addition to depleting the condensate density, the wave
function can also be excited out of the ground-state manifold to
accommodate the order-parameter singularity and form a filled defect
core.  Here we numerically find the superfluid cores of singular
vortices when the atoms exhibit a long-range and anisotropic magnetic DI.

The DI gives rise to a new spin-dependent healing length, adding to the
hierarchy of characteristic length scales arising form the contact interaction
to determine the structure of singular-vortex
cores~\cite{ruostekoski_prl_2003,lovegrove_pra_2012,borgh_prl_2016}. We analyze
the interplay of these length scales and demonstrate how the size of a
singular-vortex core is determined by the shorter of the spin-dependent healing
lengths when the DI and contact interaction both restrict breaking of the
ground-state spin condition (the ground-state phase of the bulk superfluid),
e.g., in the ferromagnetic (FM) spin-1 BEC.  On the other hand, when the contact
interaction and DI compete, such as in the spin-1 polar phase, we explain how a
singular-vortex core expands with increasing DI, beyond the size in its absence.
In addition, the anisotropy of the interaction leads to an internal spin texture
that is potentially observable in a spin-3 $^{52}$Cr condensate.

We further analyze manifestations of the interaction
anisotropy in the spin-1 model system and and show it leads to a
ground-state spin vortex that appears perpendicularly to the axis of a
slow rotation. The structure is then the result of interplay between
dipolar spin alignment and rotation as the vortex line bends to adapt
to the latter. At more rapid rotation, we demonstrate a nontrivial
interaction dependence of the angular momentum carried by a
ground-state coreless vortex. We show how it may be understood from a
competition between dipolar spin alignment and the adaptation of the
spin texture to rotation.

Drastically different spin-ordering effects can
appear in the presence of a sufficiently strong external magnetic
field, such that the DI may be averaged over the rapid spin
precession~\cite{kawaguchi_prl_2007}.  We show how the resulting
modified interaction anisotropy leads to a symmetry-breaking
core deformation with increasing DI in a stable singular vortex.
At sufficiently strong DI, the vortex core deforms into a domain wall
separating regions with opposite spin polarization.

\section{Mean-field theory of the dipolar BEC}
\label{sec:mft}

We treat the spinor BEC in the classical Gross-Pitaevskii mean-field theory,
which can be straightforwardly extended to include DI between the atoms. Here we
first give a brief overview of the salient points (for full details see, e.g.,
Ref.~\cite{kawaguchi_physrep_2012}) in the spin-1 case, and then show how the
theory is modified for spin-3 atoms.

\subsection{Spin-1}

The spin-1
condensate wave function $\Psi$ may be expressed in terms of the
atomic density and a normalized three-component spinor as
\begin{equation}
  \Psi(\rr) = \sqrt{n(\rr)}\zeta(\rr)
  = \sqrt{n(\rr)}\threevec{\zeta_+}{\zeta_0}{\zeta_-},
  \quad
  \zeta^\dagger\zeta = 1.
\end{equation}
The expectation value $\expF =
\zeta^\dagger_{\alpha}\mathbf{\hat{F}}_{\alpha\beta}\zeta_{\beta}$ of
the spin operator, defined as the vector of spin-1 Pauli matrices,
gives the condensate spin.
This relates to the magnetic dipole moment arising from the intrinsic
angular momentum of the atom as $\m =
-g_F\mu_{\mathrm{B}}\expF$~\cite{yi_prl_2004}, where $g_F$ is the
Land\'{e} factor and $\mu_{\mathrm{B}}$ is the Bohr magneton.
The Hamiltonian density including the DI is then given by
\begin{equation}
  \label{eq:hamiltonian-density}
    {\cal H} =
    h_0
    + \frac{c_0}{2}n^2
    + \frac{c_2}{2}n^2|\expF|^2
    + \frac{\cdd}{2} \int D(\rr,\rrp)\,d^3r^\prime,
\end{equation}
where
\begin{equation}
  \label{eq:dipolar-energy-density}
  D(\rr,\rrp) =
  \frac{\F(\rr)\cdot\F(\rrp)
    - 3[\F(\rr)\cdot\nhat][\F(\rrp)\cdot\nhat]}
       {|\rr-\rrp|^3}
\end{equation}
describes the interaction of dipoles at $\rr$ and $\rrp$ given by the
local condensate spin with the coupling constant $\cdd =
\mu_0\mu_{\mathrm{B}}^2g_F^2/(4\pi)$~\cite{yi_prl_2004}, where $\mu_0$
is the vacuum permeability.
We define $\F=n\expF$ and
denote the unit vector along $\rr-\rrp$ by $\nhat$.
The single-particle Hamiltonian density
\begin{equation}
  \label{eq:h0}
  h_0=\frac{\hbar^2}{2M}\abs{\nabla\Psi}^2 + \frac{1}{2}M\omega^2r^2n,
\end{equation}
where $M$ is the atomic mass,
includes the external trapping potential, which we
take to be an isotropic harmonic oscillator with frequency $\omega$.
The contact-interaction strengths are given by the scattering lengths
$a_f$ in the spin-$f$ channel of colliding spin-1 atoms as
$c_0=4\pi\hbar^2(2a_2+a_0)/(3M)+c_0^{\mathrm{d}}$ and
$c_2=4\pi\hbar^2(a_2-a_0)/(3M)+c_2^{\mathrm{d}}$. Here we have
made it explicit that the coupling constants may be
modified by contributions $c_{0,2}^{\mathrm{d}}$ arising from an absorbed
contact-interaction part of the DI (see below and Appendix~\ref{app:ft}).

The spin-1 BEC exhibits two ground state phases: a FM phase
that maximizes the condensate spin $\absF=1$ and a polar phase where
$\absF=0$ in a uniform system.  Without DI, the ground-state
phase is determined by the sign of $c_2$ with a negative value
favoring the FM phase. When magnetic DI, where the
dipole moment is proportional to the condensate spin, is present, the ground
state depends also on $\cdd$. In particular, from
Eq.~\eqref{eq:dipolar-energy-density} we can see that the DI is
minimized when spins $\absF=1$ are aligned head-to-tail.  The DI will
therefore also favor formation of a FM phase, and sufficiently large
$\cdd$ may overcome also a positive $c_2$~\cite{yi_prl_2006}.

From Eq.~\eqref{eq:hamiltonian-density} the familiar coupled Gross-Pitaevskii
equations (GPEs) describing the condensate dynamics may be derived.
Following Ref.~\cite{kawaguchi_physrep_2012}, we write the
contribution from the DI term in the equation for the
$\psi_m=\sqrt{n}\zeta_m$ spinor component as
\begin{equation}
  \label{eq:dipolar-gpe}
  i\hbar\pder{\psi_m(\rr)}{t} = \ldots
   + \cdd\sum_j\hat{\mathbf{F}}_{mj}\psi_j(\rr)\cdot\bb(\rr).
\end{equation}
The vector $\bb$ is given by
\begin{equation}
  \label{eq:b-vector-def}
    \bb(\rr)
       = \int \frac{\F(\rrp)-3\nhat[\F(\rrp)\cdot\nhat]}{|\rr-\rrp|^3}\,
       d^3r^\prime\\
\end{equation}
and is related to the magnetic field
\begin{equation}
  \label{eq:dipole-field-condensate}
  \begin{split}
    &\B(\rr) = \frac{\mu_0\{3\nhat[\m(\rrp)\cdot\nhat]-\m(\rrp)\}}
                    {4\pi |\rr-\rrp|^3}
    + \frac{2\mu_0}{3}\m(\rrp)\delta(\rr-\rrp)\\
    &= -g_F\muB\mu_0\left\{\frac{3\nhat[\F(\rrp)\cdot\nhat]-\F(\rrp)}
                                {|\rr-\rrp|^3}
    + \frac{2}{3}\F(\rrp)\delta(\rr-\rrp)\right\}
    \end{split}
\end{equation}
at $\rr$ arising from the condensate dipole moment at $\rrp$.
The factor $g_F\muB\mu_0/(4\pi)$ enters the coupling constant $\cdd$
and the $\delta$-function term yields a contact-interaction
contribution that is absorbed in $c_{0,2}$ as above (see also Appendix~\ref{app:ft}).
Integrating the remaining term over $\rrp$ yields $\bb$.

The DI term in Eq.~\eqref{eq:dipolar-gpe} is nonlocal, and its
evaluation involves finding the integral over
$\rrp$, which is computationally expensive.  However, since the
integral in Eq.~\eqref{eq:b-vector-def} has the
form of a convolution, it can be computed efficiently in Fourier
space, where the convolution of two functions becomes a
multiplication of their Fourier transforms.
For our computations we follow the formalism of
Ref.~\cite{kawaguchi_physrep_2012}, rewriting $\bb$ as
\begin{equation}
  \label{eq:b-vector}
  \bcomp_\alpha = -\sum_\beta \int
    \Qtens_{\alpha\beta}(\rr-\rrp) F_\beta(\rrp)\,
    d^3r^\prime,
\end{equation}
where the tensor $\Qtens$ is defined as
\begin{equation}
  \label{eq:q-tensor}
  \Qtens_{\alpha\beta}(\rr) =
    \frac{3\hat{r}_\alpha\hat{r}_\beta-\delta_{\alpha\beta}}{r^3},
\end{equation}
for $\rhat = \rr/r$. In Eq.~\eqref{eq:b-vector} the convolution is
explicit and Fourier transformation immediately gives
\begin{equation}
  \label{eq:b-transf}
  \tilde{\bcomp}_\alpha(\kk) =
    - \sum_\beta \QtensF_{\alpha\beta}(\kk)\tilde{F}_\beta(\kk).
\end{equation}
To compute $\bb$ we then need the Fourier transforms on the right-hand
side, where $\tilde{\F}(\kk)$ must be found numerically, while
$\QtensF(\kk)$ can be found analytically as (see Appendix~\ref{app:ft} and
Ref.~\cite{kawaguchi_physrep_2012})
\begin{equation}
  \label{eq:q-transf}
  \QtensF_{\alpha\beta}(\kk)
  = -\frac{4\pi}{3}(3\hat{k}_\alpha\hat{k}_\beta-\delta_{\alpha\beta}),
\end{equation}
where $\khat=\kk/k$.
Note that the derivation of this Fourier transform rests on nontrivial
assumptions. We provide the details in the Appendix~\ref{app:ft}.

In practical numerical calculations using Fast Fourier Transforms, the
long-range nature of the DI can lead to aliasing problems that yield
erroneous results, and accuracy may more generally be reduced.  These
problems can be avoided or mitigated by truncating the dipolar
interaction~\cite{ronen_pra_2006,blakie_pre_2009}.
In a spherical or nearly spherical system, where computations are
performed on a grid with all sides equal, the simplest solution is to
truncate the dipolar interaction at a radius $R$, such that
$\Qtens(\rr)=0$ for $r>R$.  The Fourier transform of the truncated
interaction is then
\begin{equation}
  \label{eq:q-transf-trunc}
  \begin{split}
  &\QtensF^{r<R}_{\alpha\beta}(\kk)
  = \int_{r<R} \Qtens_{\alpha\beta}(\rr)e^{-i\kk\cdot\rr}\,d^3r \\
  &= -4\pi(3\hat{k}_\alpha\hat{k}_\beta-\delta_{\alpha\beta})
  \left(\frac{1}{3} + \frac{kR\cos(kR) - \sin(kR)}{(kR)^3}\right),
  \end{split}
\end{equation}
which is the spherical cut-off found in Ref.~\cite{ronen_pra_2006},
straightforwardly generalized to the spinor case (see Appendix~\ref{app:ft}).

In the presence of an external magnetic field
$\mathbf{B}_{\mathrm{ext}}=B_{\mathrm {ext}}\zhat$, the condensate
spin precesses with the Larmor
frequency $\omega_{\mathrm{L}}=g_F\mu_{\mathrm{B}}B_{\mathrm {ext}}/\hbar$.
In a sufficiently strong field, the
precession is rapid compared with the DI-induced spin dynamics.  It is
then convenient to describe the condensate in the spin-space
frame rotating at the Larmor frequency
through the transformation $\zeta_m(\rr,t) \to
\zeta_m(\rr,t)e^{-im\omega_{\mathrm{L}}t}$~\cite{kawaguchi_prl_2007,kawaguchi_pra_2010}. This
leaves all terms of Eq.~\eqref{eq:hamiltonian-density} invariant,
except the dipolar interaction. (Also the linear Zeeman term that
would arise from the magnetic field is canceled and we assume any
quadratic Zeeman energy to be small.)  The
modified dipolar interaction is found as a time average over the
period of the Larmor precession~\cite{kawaguchi_pra_2010}:
\begin{equation}
  \label{eq:q-tensor-l}
    \Qtens_{\alpha\beta}^\mathrm{L}(\rr)
    = \frac{3\hat{r}_z^2-1}{r^3}
      \frac{3\delta_{z\alpha}\delta_{z\beta}-\delta_{\alpha\beta}}{2}
\end{equation}
Also in this case we truncate the dipolar interaction at a radius $R$ for
computational purposes.  Its Fourier transform then becomes (see Appendix~\ref{app:ft})
\begin{equation}
  \label{eq:q-transf-l-trunc}
  \begin{split}
    \QtensF^{\mathrm{L},r<R}_{\alpha\beta}(\kk)
    = & -2\pi(\hat{k}_z^2-1)
    (3\delta_{z\alpha}\delta_{z\beta}-\delta_{\alpha\beta})\times\\
    &\left[\frac{1}{3} + \frac{kR\cos(kR) - \sin(kR)}{(kR)^3}\right].
  \end{split}
\end{equation}

\subsection{Spin-3}

The spin-1 condensate provides a useful system where the physical principles
underlying the dipolar effects in a spinor BEC can be illustrated.
Dipolar spin-1 BECs could potentially be realized using Na or Rb atoms  by suppressing the
$s$-wave scattering lengths via optical or microwave Feshbach resonances.
However,
large magnetic dipole moments are exhibited, e.g., by $^{52}$Cr, which
is a spin-3 atom.  In this case, the condensate wave function becomes
a seven-component spinor with $\zeta =
(\zeta_{+3},\dots,\zeta_{-3})^T$ and
a Hamiltonian density given
by~\cite{kawaguchi_physrep_2012,diener_prl_2006,santos_prl_2006}
\begin{equation}
  \label{eq:spin-3-hamiltonian-density}
  \begin{split}
    {\cal H} =
    &h_0
    + \frac{c_0}{2}n^2
    + \frac{c_2}{2}n^2|\expF|^2
    + \frac{c_4}{2}n^2|A_{00}|^2\\
    &+ \frac{c_6}{2}n^2\sum_{j=-2}^{+2}|A_{2j}|^2
    + \frac{\cdd}{2} \int D(\rr,\rrp)\,d^3r^\prime,
  \end{split}
\end{equation}
where two additional interaction terms, compared with
Eq.~\eqref{eq:hamiltonian-density},  appear as a result of the
$s$-wave scattering of spin-3 atoms.  These depend on the amplitudes
\begin{equation}
  \label{eq:A00}
  A_{00} = \frac{1}{\sqrt{7}} \left(2\zeta_{+3}\zeta_{-3} -
    2\zeta_{+2}\zeta_{-2} + 2\zeta_{+1}\zeta_{-1} - \zeta_0^2\right)
\end{equation}
and
\begin{equation}
  \label{eq:A2j}
  \begin{split}
    &A_{20} = \frac{1}{\sqrt{7}} \left(\frac{5}{\sqrt{3}}\zeta_{+3}\zeta_{-3}
      - \sqrt{3}\zeta_{+1}\zeta_{-1} + \sqrt{\frac{2}{3}}\zeta_0^2\right),\\
    &A_{2\pm1} = \frac{1}{\sqrt{7}}
    \left(\frac{5}{\sqrt{3}}\zeta_{\pm3}\zeta_{\mp2}
      - \sqrt{5}\zeta_{\pm2}\zeta_{\mp1}
      + \sqrt{\frac{2}{3}}\zeta_{\pm1}\zeta_0\right),\\
    &A_{2\pm2} = \frac{1}{\sqrt{7}}
    \left(\frac{10}{\sqrt{3}}\zeta_{\pm3}\zeta_{\mp1}
      - \sqrt{\frac{20}{3}}\zeta_{\pm2}\zeta_{0}
      + \sqrt{2}\zeta_{\pm1}^2\right),
  \end{split}
\end{equation}
respectively. The interaction strengths $c_{0,2,4,6}$ are found from
the scattering lengths of the four spin channels of colliding spin-3
atoms as $c_0=4\pi\hbar^2(9a_4+2a_6)/(11M)+c_0^{\mathrm{d}}$,
$c_2=4\pi\hbar^2(a_6-a_4)/(11M)+c_2^{\mathrm{d}}$,
$c_4=4\pi\hbar^2(11a_0-21a_4+10a_6)/(11M)+c_4^{\mathrm{d}}$, and
$c_6=4\pi\hbar^2(11a_2-18a_4+7a_6)/(11M)+c_6^{\mathrm{d}}$,
where the coupling constants may again be modified by contact part of the DI.
The dipolar interaction is again given by
Eq.~\eqref{eq:dipolar-energy-density}, where, the spin
operator $\mathbf{\hat{F}}$ is now the vector of $7\times7$
spin-3 Pauli matrices.  From Eq.~\eqref{eq:spin-3-hamiltonian-density}
seven coupled GPEs for the components of the spinor wave function may
be derived.  Using the spin-3 $\mathbf{\hat{F}}$ operator in
Eqs.~\eqref{eq:dipolar-energy-density} and
\eqref{eq:b-vector-def} yields the DI contribution, which can then
be treated analogously to the spin-1 case.

The spin-3 BEC exhibits a complex family of phases exhibiting different
symmetries~\cite{kawaguchi_pra_2011}. Here
we concentrate on $^{52}$Cr, where current measurements of the scattering
lengths~\cite{werner_prl_2005,pasquiou_pra_2010,de-paz_pra_2014}
predict an $A$-phase ground state with $\absF=0$ in a uniform system.
Nevertheless, the DI may influence the structure of singular vortices
as they develop superfluid cores with nonzero spin.

\section{Results}
\label{sec:results}

We now employ the mean-field theory outlined in Section~\ref{sec:mft} to study
the internal core structure of vortices. We find the vortex solutions by solving
the coupled GPEs derived from Eq.~\ref{eq:hamiltonian-density} in the frame
rotating with frequency $\Omega$ around the $z$ axis: $\mathcal{H} \to
\mathcal{H} - \Omega\langle\hat{L}_z\rangle$, where $\hat{L}_z$ is the $z$
component of the angular-momentum operator.  This is done using a successive
overrelaxation method~\cite{numerical-recipes} to find stationary solutions in
the spin-1 model, while we have used imaginary-time propagation for particular
results and to solve the spin-3 GPEs for $^{52}$Cr.

We first find our main results considering a spin-1 BEC.  While higher-spin
atoms are necessary to reach large magnetic dipole moment, the spin-1 system
provides a useful model where the physics arising from the DI can be illustrated
and compared with known results in a non-dipolar condensate.  Since the DI is
given by Eq.~\eqref{eq:dipolar-energy-density} regardless of the atomic spin and
its effects arise from generic properties of the interaction and the spinor
condensates, the physical principles illustrated by the results can be expected
to apply more broadly also in higher-spin systems. Dipolar BECs can also be
realized with weak dipolar interactions, provided that the other nonlinearities
are even weaker (see aslo Appendix~\ref{app:interaction-strengths}). We keep the
spin-independent interaction strength fixed at $Nc_0=10^4\hbar\omega\ell^3$,
where $\ell$ is the oscillator length $\ell=\sqrt{\hbar/(M\omega)}$ of the
harmonic trap and $N$ is the number of atoms in the condensate.  We allow $c_2$
to vary around $c_0/c_2\simeq-216$, which corresponds to $^{87}$Rb, the most
commonly used atom with FM interactions in spin-1 experiments. We then study how
the vortex structure varies with $\cdd$.

We further briefly consider a polar BEC with $c_0/c_2\simeq28$, corresponding to
$^{23}$Na. We then find the stable core structure of singular half-quantum
vortices in a spin-3 $^{52}$Cr BEC with with and without the corresponding DI,
and analyze these in light of the spin-1 model.

\subsection{Weak magnetic field}

\subsubsection{Spin vortex}

For our spin-1 model, we first consider a condensate in the FM
interaction
regime, $c_2<0$, such that in a uniform system $\absF$ is maximized.
In this case, different FM spinors are related by three-dimensional
rotations of the orthonormal triad formed by $\expF$ and two vectors
perpendicular to it.  The order-parameter space is therefore $\SO(3)$,
which supports only two topologically distinct classes of
vortices~\cite{kawaguchi_physrep_2012}: nonsingular coreless
vortices and singly quantized singular vortices.  Here we first
consider the singular vortex, whose core structure in the absence of
DI we studied in detail in
Refs.~\cite{lovegrove_pra_2012,lovegrove_pra_2016}.

In a nondipolar condensate, the stabilization of a vortex line usually
requires a sufficiently rapid external rotation.  While the ground
state in a rotating FM spin-1 condensate is generally predicted to be
made up of coreless
vortices~\cite{mizushima_prl_2002,martikainen_pra_2002}, a singular
FM vortex can also be energetically (meta-)stable for a range of
trap-rotation frequencies~\cite{lovegrove_pra_2012}, and is predicted
to form the ground state when the coreless vortex is destabilized by
conservation of a weak
magnetization~\cite{lovegrove_prl_2014,lovegrove_pra_2016}.
Even though spin vortices that carry no mass circulation can form in a
spinor BEC, one would not generally expect them to be energetically
stable. In a FM spin-1 BEC, a spin vortex can be stabilized by
magnetic fields, e.g., in a Ioffe-Pritchard
trap~\cite{bulgakov_prl_2003}, and a BEC with FM interactions
initially in a polar state can be dynamically unstable towards
spin-vortex formation~\cite{saito_prl_2006}.
In the FM phase, mass circulation alone
is not quantized and a spin
vortex can continuously pick up angular momentum through local spin
rotations to stabilize it in a rotating
trap~\cite{mizushima_pra_2002,lovegrove_pra_2012}.

\begin{figure}[tb]
  \centering
  \includegraphics[width=\columnwidth]{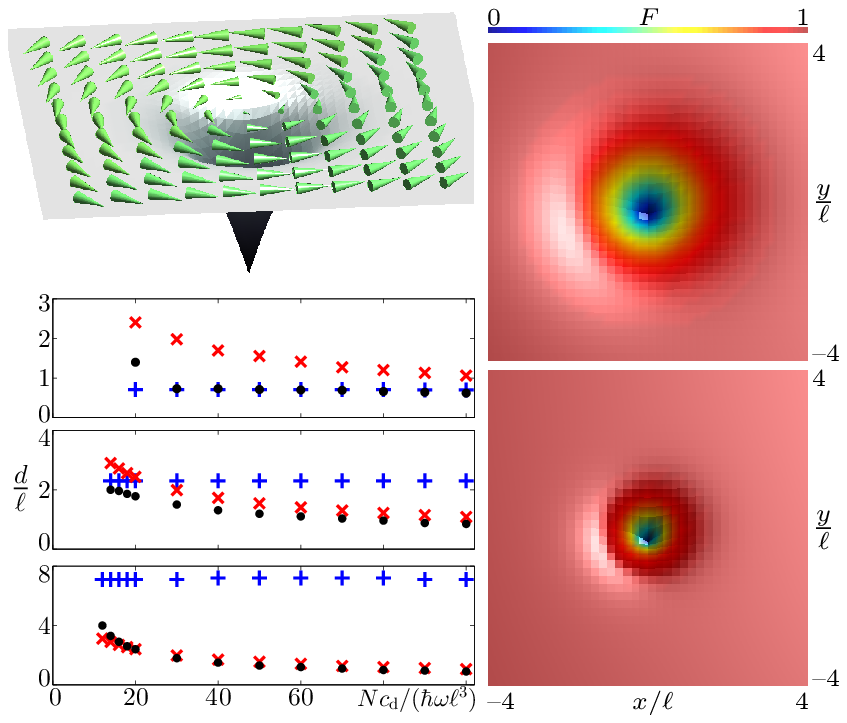}
  \caption{Top left: Spin structure of a spin vortex in a nonrotating
    condensate. Surface plot indicates $\absF$, showing the vortex
    core. Cones indicate $\expF$, exhibiting a tangential disgyration
    that minimizes the DI energy. Bottom left:
    Size of the vortex core ($\bullet$)
    compared with $\xi_F$ ($+$) and $\xidd^\prime$ ($\times$) as
    functions of $\cdd$ (see also Appendix~\ref{app:interaction-strengths}).
    Subpanels from top to bottom: $Nc_2 = -463\hbar\omega\ell^3$,
    $-46.3\hbar\omega\ell^3$, and $-4.63\hbar\omega\ell^3$ ($Nc_0 =
    10^4\hbar\omega\ell^3$). The spin vortex is stable above a
    critical $\cdd$ that depends on $c_2$.  The core size is then
    well predicted by the smaller of the two healing lengths.
    Right: Surface plot of $F=\absF$ (color scale) showing the different
    sizes of the vortex core for $N\cdd =
    16\hbar\omega\ell^3$ (top) and $N\cdd = 100\hbar\omega\ell^3$
    (bottom) for $Nc_2=-46.3\hbar\omega\ell^3$.
  }
  \label{fig:spin-vortex-core}
\end{figure}

When a magnetic DI is present,
however, the situation changes due to the anisotropy of the interaction,
which strives to arrange the dipoles in a head-to-tail configuration
that minimizes the interaction energy.  Beyond a critical $\cdd$, it
then becomes possible for a singular spin
vortex carrying no circulation to form in the ground state even in a
nonrotating
condensate~\cite{yi_prl_2006,kawaguchi_prl_2006b,takahashi_prl_2007}. The
structure of the stable spin vortex is shown in
Fig.~\ref{fig:spin-vortex-core}.

The tangential disgyration exhibited by the condensate spin is a
consequence of the DI.  The singular FM vortex in the spin-1 BEC can
exhibit a wide range of associated spin textures that can be
transformed into each other through local and continuous
transformations.  In addition to the tangential disgyration shown in
Fig.~\ref{fig:spin-vortex-core}, radial and cross disgyrations are
possible, as well as asymptotically uniform spin textures. In the
absence of DI, these different spin structures are energetically (near)
degenerate~\cite{lovegrove_pra_2012}.  Here, this degeneracy is broken
by the directional dependence of the DI.  The tangential disgyration
corresponds to the greatest head-to-tail alignment of the spins, and
therefore minimizes the DI energy, energetically locking in the spin
texture. The vortex can then be described by the spinor wave function
\begin{equation}
  \label{eq:spin-vortex}
  \zeta = \frac{i}{\sqrt{2}}
          \threevec{-\sqrt{2}e^{-i\phi}\cos^2\frac{\beta}{2}}
                   {\sin\beta}
                   {\sqrt{2}e^{i\phi}\sin^2\frac{\beta}{2}},
\end{equation}
where $\phi$ is the azimuthal angel and $F_z=\cos\beta$. For
$\beta=\pi/2$, such that $\expF$ lies in the $xy$ plane as in
Fig.~\ref{fig:spin-vortex-core}, the vortex is a pure spin vortex that
carries no angular momentum.

In a scalar BEC, the superfluid density vanishes on the line
singularity of the order parameter that constitutes a vortex line.
In the multicomponent
order parameter of a spinor BEC, by contrast, a vortex-line
singularity can also be accommodated by exciting the wave
function out of its ground-state manifold.  In a spin-1 BEC, the
resulting filled vortex core becomes energetically favorable when
$c_2$ is small compared with $c_0$, which is usually the case.  This
can be understood from the healing lengths arising from the
contact-interaction terms.  These are the density and spin healing lengths
$\xi_n=\hbar/(2Mc_0n)^{1/2}$ and $\xi_F=\hbar/(2M|c_2|n)^{1/2}$ that
describe the characteristic length scales of deviations from the
corresponding ground-state condition.  By breaking the spin condition
instead of depleting the density, the defect core can expand to the
larger healing length and lower its
energy~\cite{ruostekoski_prl_2003,lovegrove_pra_2012}.  A singular FM
vortex then develops a superfluid core exhibiting the polar phase on
the line singularity.  When the vortex is represented by
Eq.~\eqref{eq:spin-vortex}, this corresponds to the $\zeta_0$
component occupying the singular lines in $\zeta_\pm$.

In the dipolar spinor BEC considered here, an additional interaction
term appears in the Hamiltonian density,
Eq.~\eqref{eq:hamiltonian-density}.  Unlike the interaction terms
arising from the $s$-wave scattering, the DI term is nonlocal.
However, it is still possible to associate with it a length scale
\begin{equation}
  \label{eq:xidd}
  \xidd = \frac{\hbar}{\sqrt{2M|\cdd|n}}.
\end{equation}
It was shown in Ref.~\cite{kawaguchi_prl_2006b} that dipole-induced
spin textures such as the ground-state spin vortex can form when the
extent of the condensate exceeds $\xidd$.
Here we show
that the dipolar healing length also interacts nontrivially with the
other characteristic length scales to affect the vortex-core structure in
the dipolar spinor condensate.

Specifically, we find that the dipolar healing length becomes important
for the structure of a singular-vortex core when it is \emph{shorter}
than the spin healing length: $\xidd\lesssim\xi_F$.  This is contrary
to the nondipolar case, where the core structure of
a singular vortex is determined by
the largest healing length~\cite{lovegrove_pra_2012}.
The healing lengths in that case are associated with different and
independent ground state conditions (superfluid density and spin
magnitude). In the case of magnetic DI, however,
$\xi_F$ and $\xidd$ both relate to the condensate
spin.  In particular, when $c_2<0$ the contact interaction and the DI
both energetically favor $\absF=1$.
Any perturbation of $\absF$ must then heal back to the bulk value over
the shortest of the spin-dependent healing lengths.
Consequently, the size of the core becomes dependent on $\cdd$ when
$\xidd$ becomes comparable to $\xi_F$ as illustrated in
Fig.~\ref{fig:spin-vortex-core}.

Unlike the contact interaction, the effective strength of the DI depends on the
relative orientation of the dipoles.  In a head-to-tail arrangement, the
effective strength of the interaction is $-2\cdd$ [cf.\
Eq.~\eqref{eq:dipolar-energy-density} for $\F(\rrp)=\F(\rr)$].  In the context
of the spin vortex shown in Fig.~\ref{fig:spin-vortex-core}, $\xi_F$ should
therefore be compared with $\xidd^\prime \equiv \xidd/\sqrt{2}$. In the bottom
left panels of Fig.~\ref{fig:spin-vortex-core}, we plot both $\xi_F$ and
$\xidd^\prime$, together with the vortex core size (defined as the diameter of
the core at $\absF=1-e^{-1}$), as functions of $\cdd$.   [For simplicity we here
treat $\cdd$ as a free parameter within the spin-1 model system.
Appendix~\ref{app:interaction-strengths} outlines how the dimensionless
nonlinearity $N\cdd/(\hbar\omega\ell^3)$ can be varied also for fixed $\cdd$,
corresponding to some particular magnetic dipole moment.] The middle subpanel
corresponds to $Nc_2=-46.3\hbar\omega\ell^3$ (corresponding to $^{87}$Rb, whose
physical dipole moment also gives $N\cdd\simeq4.2\hbar\omega\ell^3$, for
comparison), while in the top and bottom panels, $c_2$ is one order of magnitude
stronger and weaker, respectively. For the strong $c_2$, $\xi_F<\xidd^\prime$
over the range of the plot, and the core size remains nearly unaffected by
$\cdd$ and is well predicted by $\xi_F$ (except at the very onset of stability).
Conversely, for the weak $c_2$, $\xi_F>\xidd^\prime$, with the latter quantity
corresponding well to the core size.  In the middle panel, $\xi_F \simeq
\xidd^\prime$ and the two cross as $\cdd$ increases.  For large $\cdd$ the core
size follows $\xidd^\prime$, while at small $\cdd$ the influence of the now
smaller $\xi_F$ becomes evident. These fairly simple principles then
characterize the behavior of a singular-defect core as DI is varied.

\begin{figure}[tb]
  \centering
  \includegraphics[width=\columnwidth]{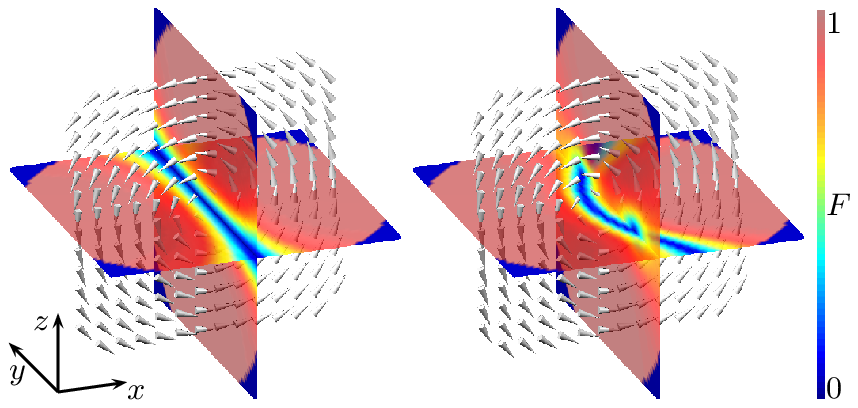}
  \caption{Spin vortex in nonrotating (left) and slowly rotating,
    $\Omega=0.10\omega$ (right), trap. Color map shows $F=\absF$
    highlighting the polar vortex core, while cones show the spin vector
    in the $y=0$ plane. In the rotating system, the
    vortex line forms perpendicular to the rotation axis.
    $Nc_0=10^4\hbar\omega\ell^3$, $Nc_2=-46.3\hbar\omega\ell^3$,
    $N\cdd=50\hbar\omega\ell^3$.
  }
  \label{fig:perp-spin-vortex}
\end{figure}

For a nonrotating cloud in a 3D isotropic trap, there is no preferred
direction for the spin-vortex line (in the absence of Zeeman shifts).  In
Fig.~\ref{fig:spin-vortex-core}, the vortex line
coincides with the $z$ axis, while the left panel of
Fig.~\ref{fig:perp-spin-vortex} shows a spin vortex line in the $xy$
plane.  Considering now a slowly rotating trap, the axis of
rotation represents a preferred spatial direction.  Vortices
stabilized by rotation would then form parallel to the rotation axis,
as in the nondipolar case~\cite{lovegrove_pra_2012}.
In a highly oblate dipolar spinor condensate, the spin vortex
with axial symmetry around the $z$ direction persists
up to a critical rotation frequency~\cite{simula_jpsj_2011}.
For our isotropic trap, however, we find that the ground state in a
slowly rotating
condensate exhibits a spin vortex forming perpendicularly to the
rotation axis when $\cdd$ is sufficiently large
(Fig.~\ref{fig:perp-spin-vortex}). While a solution with a spin vortex
parallel to the rotation also exists, this has a higher energy.
The orientation of the vortex line shows that similarly to the
nonrotating case, the vortex forms due to minimization of the DI
energy, rather than because of the rotation.  The effect of the
rotation is instead to increasingly bend the vortex line, as
illustrated in Fig.~\ref{fig:perp-spin-vortex}.

\subsubsection{Coreless vortex}

When the condensate rotates sufficiently rapidly, we find that the
ground state is a coreless vortex along the rotation axis.  This was
also found to be the case in the highly oblate trap in
Ref.~\cite{simula_jpsj_2011}. Coreless vortices are also predicted to
make up the ground state in a rotating nondipolar FM
condensate~\cite{mizushima_prl_2002,martikainen_pra_2002}, unless
destabilized through conservation of a sufficiently weak
magnetization~\cite{lovegrove_prl_2014}.
The coreless vortex is characterized by a
nonsingular spin texture in which the superfluid circulation varies
continuously as the spin bends from $\expF=\zhat$ on the vortex line
toward its asymptotic direction.
The boundary condition on the spin texture away from the vortex line
is however not fixed, allowing it to adapt to the imposed rotation,
bending more sharply towards the $-\zhat$ direction as rotation increases.

\begin{figure}[tb]
  \centering
  \includegraphics[width=\columnwidth]{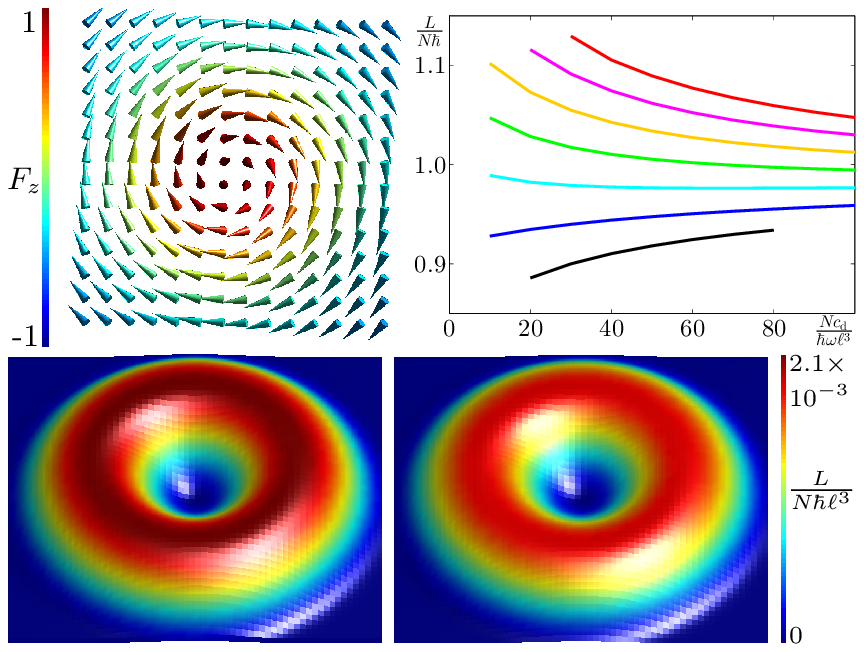}
  \caption{Top left: Spin texture of the coreless vortex for
    $\Omega=0.175\omega$ and $N\cdd=10\hbar\omega\ell^3$. Cones show the spin
    vector in the $x,y$ plane, perpendicular to the vortex line, with the color
    scale indicating the $z$ component. The DI
    causes the condensate spin to bend toward a tangential disgyration
    away from the vortex line.  Top right: Total angular momentum
    carried by the coreless vortex as a function of $\cdd$.  The lines
    from bottom to top correspond to trap rotation $\Omega=0.14\omega$
    through $\Omega=0.20\omega$ in steps of $0.01\omega$. The varying
    dependence on $\cdd$ and the convergent behavior at large values
    is the result of the spin texture simultaneously adapting to DI and
    imposed rotation.  Bottom: Angular-momentum density distribution
    for $\Omega=0.18\omega$ and $N\cdd=10\hbar\omega\ell^3$ (left) and
    $100\hbar\omega\ell^3$ (right). The panels show the same $13\ell$ by
    $13\ell$ cut-out and use the same color scale. In all panels
    $Nc_0=10^4\hbar\omega\ell^3$, $Nc_2=-46.3\hbar\omega\ell^3$}
  \label{fig:cl}
\end{figure}

The DI introduces a competing mechanism that strives to align the spins
in the head-to-tail configuration of a tangential disgyration similar
to the spin vortex.  This leads
to the formation of a coreless vortex with the spin texture
shown in Fig.~\ref{fig:cl}, where the spin vector bends gradually
into the tangential disgyration.  Far from the vortex line, the DI
thus determines the spin texture by the same mechanism as for the spin
vortex.  Note, however, that while the two vortices appear
superficially similar, both exhibiting tangential disgyrations of the
spin vector at large length scales, their topology is
distinctly different.  This is easily established from the complex
phases of the individual spinor components, which exhibit
$(2\pi,0,-2\pi)$ winding in the singular spin vortex, but
wind by $(0,2\pi,4\pi)$ in the coreless vortex.  In the latter case,
the vortex can be removed through purely local spin rotations, provided that
the value of the spin is free to rotate at the edge of the cloud.

The structure of the coreless vortex can be understood as the result
of competition between rotation and DI.
While the spin texture strives to adapt to the imposed rotation
asymptotically forming an angle with the
$z$ axis that depends on the rotation frequency, the DI strives to
align the asymptotic texture in the $xy$ plane.  This competition is
reflected in the total angular momentum $L$ carried by the vortex, which
becomes dependent on $\cdd$ at fixed trap rotation.  When the trap
rotates slowly, the rotation alone is not enough to bring the
asymptotic texture into the $xy$ plane.  Increasing $\cdd$ will then
result in a more sharply bending texture that carries additional
angular momentum, such that $L$ increases as a function of $\cdd$.
On the other hand, a rapid rotation causes the asymptotic
spin texture to acquire a negative $F_z$ component in order to provide
sufficient circulation. Increasing $\cdd$ then has the opposite
effect, causing the spin to bend less sharply in order to bring it
more in line with the $xy$ plane.  This causes $L$ to decrease
with with DI strength.
Consequently, as the DI becomes more dominant
with increasing $\cdd$, $L$ becomes less sensitive to the trap
rotation frequency.

\subsubsection{Polar condensate}
\label{sec:polar}

The DI couples to the condensate spin through
Eq.~\eqref{eq:dipolar-energy-density}.  So far we have explored how
this leads to consequences for the formation, stability and structure
of in the FM phase, where $\absF$ is maximized in the bulk condensate.
In the polar phase of the spin-1 BEC, $\absF=0$ in a uniform system.
However, when a singular vortex is present, nonzero $\absF$ can appear
in the defect core, and DI can still affect its structure.  The
physics, however, exhibits differences from the FM case, where the
contact and dipolar interactions both strive to maximize the
condensate spin.  Here, by contrast, the
interactions compete, with the contact interaction favoring
$\absF=0$.  The size of a superfluid core with nonzero spin is then
not limited by the dipolar interaction, which now favors breaking of
the ground-state spin condition.  One may then expect the
presence of the DI to lead to an enlarged core, and our numerical
simulations confirm these simple principles.

We explore this by considering a stable singular half-quantum vortex in a
rotating system. We keep $c_0$ the same as in the FM examples, but now take
$c_0/c_2\simeq28$ corresponding to $^{23}$Na. Energy relaxation causes the
vortex to develop a superfluid core that breaks the ground-state spin condition,
reaching the FM phase on the singular line~\cite{lovegrove_pra_2016}. In the
absence of DI the size of the vortex core is determined by the spin healing
length $\xi_F$. We find that as the strength of the DI increases from $\cdd = 0$
to $c_0/\cdd = 200$, the size of the vortex core increases by $\sim35\%$ for
$Nc_0=10^4\hbar\omega\ell^3$ and rotation frequency in the range $0.12\omega
\leq \Omega \leq 0.17\omega$.

\subsection{Precession-averaged dipolar interaction in a magnetic
  field}

\begin{figure}[tb]
  \centering
  \includegraphics[width=\columnwidth]{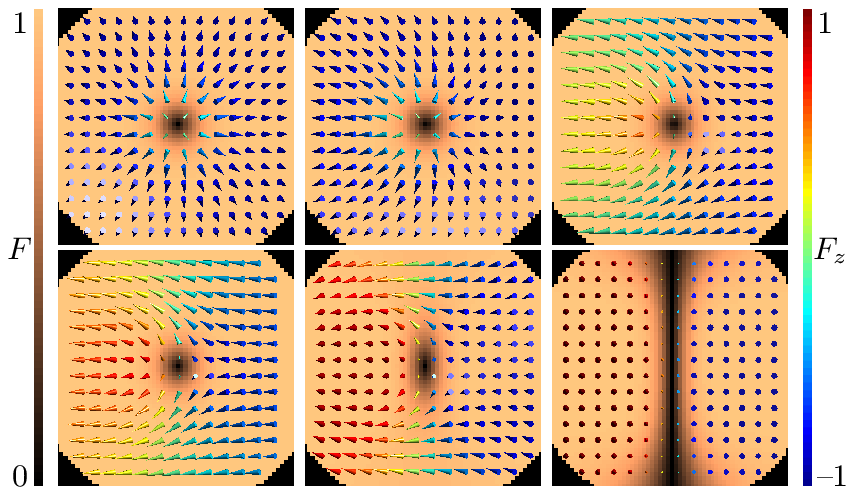}
  \caption{Structure of the stable singular-vortex core ($z=0$ cross section,
    perpendicular to the vortex line) as the
    effective DI averaged over the Larmor precession increases. Color
    map shows $F=\absF$, while cones show the spin vector $\expF$
    (color indicates $F_z$).  From top left to bottom
    right $N\cdd = 0, 6, 8, 10, 50$ and $100\hbar\omega\ell^3$. The core
    deforms by breaking axial symmetry to form an ellipsoidal cross section.
    Eventually the core covers the diameter of the condensate and
    forms a domain wall between regions of $F_z=\pm1$. In all panels
    $Nc_0=10^4\hbar\omega\ell^3$, $Nc_2=-46.3\hbar\omega\ell^3$, and
    $\Omega=0.13\omega$.
  }
  \label{fig:larmor-singular}
\end{figure}

In experiment, the condensate may be subject to residual or deliberately imposed
external magnetic fields.  In the presence of the magnetic field, the condensate
spin exhibits precession around the field direction at the Larmor frequency
$\omega_{\mathrm{L}}$.  If the field is sufficiently strong, the Larmor
precession will be rapid compared with the spin dynamics resulting from the DI.
On the latter time scale, the condensate then experiences an effective DI that
corresponds to the averaging of the bare DI over the period of the Larmor
precession. The resulting reduced DI is given by Eq.~\eqref{eq:q-tensor-l}. This
removes some of the anisotropy of the bare DI and therefore leads to degeneracy
between some spin configurations that would otherwise have different energies.
Here we show that this can have a profound effect on the spin structure of
singular vortices. Figure~\ref{fig:larmor-singular} shows the vortex core and
spin texture of a stable singular vortex in a rotating system. For sufficiently
small $\cdd$, the vortex is axially symmetric, exhibiting a radial disgyration
in the $xy$ components of the spin vector.

Increasing $\cdd$ results in a deformation of the
vortex core, breaking the axial symmetry.  The spins rotate toward a
configuration where the asymptotic spin texture exhibits a nearly
uniform projection onto the $xy$ plane, while $F_z$ bends across the
condensate. At the same time, the polar core of the vortex deforms to
exhibit an elliptical cross section whose eccentricity increases with
$\cdd$.  Eventually the deformation of the vortex core becomes large
enough that its extent along the major axis of the ellipse reaches the
condensate size.  As shown in Fig.~\ref{fig:larmor-singular} the
condensate the exhibits a
polar domain wall separating two halves of the cloud with oppositely
aligned spin $\expF=\pm\zhat$.  In the isotropic spinor condensate,
there is no analog of this deformation when the DI cannot be averaged
over the Larmor precession period.  However, a similar anisotropic
deformation has been predicted in a two-dimensional, scalar BEC with
fixed dipole moments~\cite{mulkerin_prl_2013}

\subsection{Singular vortex in a spin-3 $^{52}$Cr BEC}
\label{sec:spin-3-results}

\begin{figure}[tb]
  \centering
  \includegraphics[width=\columnwidth]{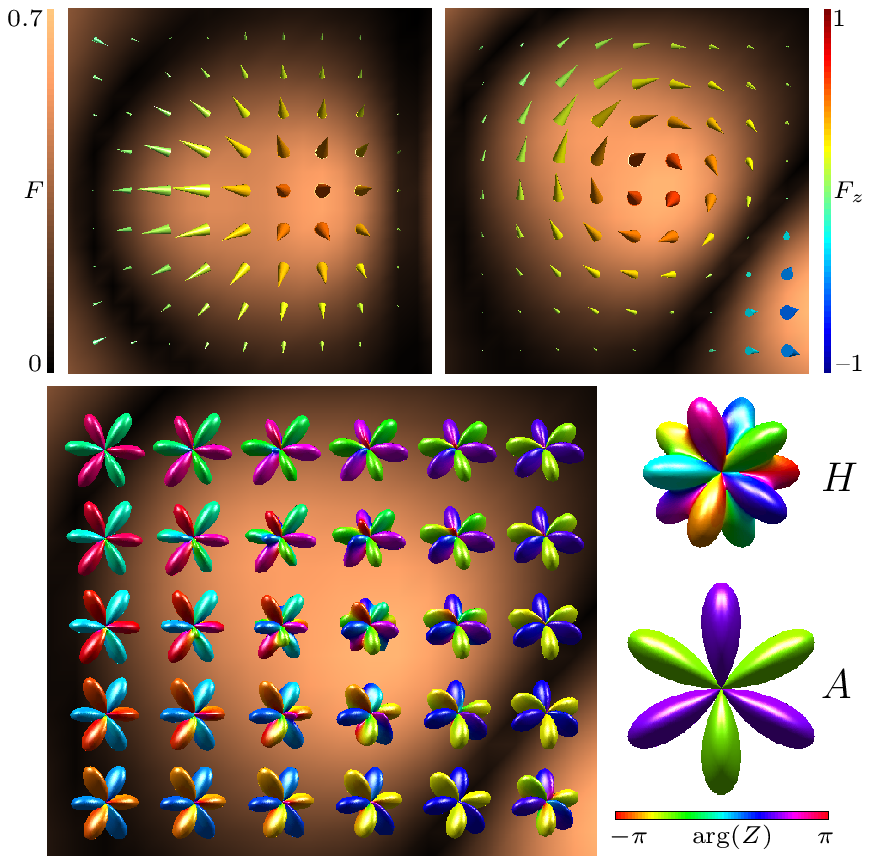}
  \caption{Top: Core structure of a singular half-quantum vortex in the
    $A$-phase of a spin-3 BEC in the absence (left) and presence
    (right) of DI corresponding to $^{52}$Cr. Color
    map shows $F=\absF$, while cones show the spin vector $\expF$
    (color indicates $F_z$). The panels show a
    $2.4\ell\times2.4\ell$ region around the vortex line, which is stable
    together with a second half-quantum vortex (not shown) in a system
    rotating with frequency $\Omega=0.43\omega$. The contact
    interaction corresponds to $^{52}$Cr with
    $Nc_0=10^3\hbar\omega\ell^3$.
    Bottom left: Change of the order parameter symmetry showing
    the transition from the $A$-phase bulk towards the $H$-phase in
    the core when DI is present (background color map shows $F$ for
    reference). Bottom right: Spherical-harmonics representations of
    the $A$- and $H$-phase order parameters for reference.
  }
  \label{fig:spin3}
\end{figure}

The spin-1 BEC provides a good model system for theoretically exploring the
physical principles of DI effects on vortices in spinor BECs.
However, a stronger
dipole moment of $6\muB$ is found in $^{52}$Cr,
which can be used to create a spin-3
condensate~\cite{diener_prl_2006,santos_prl_2006}.  Measurements of
the $s$-wave scattering
lengths~\cite{werner_prl_2005,pasquiou_pra_2010,de-paz_pra_2014} yield
$c_0/c_2 \simeq 20$, $c_0/c_4 \simeq -4.6$, $c_0/c_6 \simeq -1.5$ for the
contact-interaction strengths in
Eq.~\eqref{eq:spin-3-hamiltonian-density}, while $c_0/\cdd \simeq 177$.
The ground-state determined by the
$s$-wave interaction is then the so-called $A$-phase in a uniform
system~\cite{kawaguchi_pra_2011}, with a
representative order-parameter
$\zeta=(1/\sqrt{2},0,0,0,0,0,1/\sqrt{2})^T.$
This exhibits $\absF=0$ and the order parameter has the discrete
hexagonal symmetry of the dihedral-6 group
$D_6$~\cite{yip_pra_2007,barnett_pra_2007,kawaguchi_pra_2011}, illustrated in
Fig.~\ref{fig:spin3} using the spherical-harmonics representation
\begin{equation}
  \label{eq:spherical-harmonics}
  Z(\theta,\phi)=\sum_{m=-3}^{+3}Y_{3,m}(\theta,\phi)\zeta_m.
\end{equation}
In a biaxial-nematic spin-2 BEC, the related but simpler
dihedral-4 point-group symmetry already leads to highly complex core
structures of a half-quantum vortex~\cite{borgh_prl_2016}.  As a
result of the $D_6$ symmetry, the spin-3 $A$-phase vortices are
also non-Abelian (i.e., the different topological charges do not all
commute), leading to the restricted reconnection dynamics of
vortices also predicted in the cyclic and biaxial-nematic spin-2
phases~\cite{kobayashi_prl_2009,borgh_prl_2016}.

The $A$-phase $D_6$ order parameter supports a half-quantum vortex where the
$\pi$ winding of the condensate phase is compensated by a $\pi/3$ spin rotation.
This is the simplest vortex that carries angular momentum and could therefore be
stabilized by rotation. Figure~\ref{fig:spin3} shows the relaxed core structure
of one out of a pair of such half-quantum vortices in a condensate with and
without the DI corresponding to $^{52}$Cr, for the case where any external
magnetic field is negligible (i.e., the spin precession is not assumed to be
rapid). We find that energy relaxation leads to the condensate approaching the
$H$-phase~\cite{kawaguchi_pra_2011} on the singular line. A representative
$H$-phase order parameter can be written
$\zeta=(\sqrt{(2+F)/5},0,0,0,0,\sqrt{(3-F)/5},0)^T$, where $F=\absF$. This phase
exhibits a five-fold rotational symmetry, shown in Fig.~\ref{fig:spin3}.  The
bottom left panel also uses the spherical-harmonics representation,
Eq.~\eqref{eq:spherical-harmonics}, to illustrate the change of the order
parameter form the bulk $A$-phase to the vortex core. The $H$-phase further
exhibits a parameter-dependent condensate spin magnitude that is determined by
energy relaxation. Here we find $\absF>0$ in the vortex core, which therefore
breaks the ground-state spin condition. The effects of DI are then similar to
the polar half-quantum vortex with FM core in the spin-1 model
(section~\ref{sec:polar}), where increasing DI leads to an increase in core size
since the DI favors the nonzero spin.  In the spin-3 vortex, however, $\absF$ is
not restricted to a particular value on the vortex line, but is determined by
energy relaxation and depends on the interaction parameters. Comparing the
stable vortex cores, we find a slightly increased spin magnitude in the presence
of DI, illustrating the general principle that was also demonstrated for the
spin-1 dipolar BEC.

However, the presence of the DI is also reflected in the internal spin texture
of the vortex core,
which is reminiscent of the coreless vortex in
Fig.~\ref{fig:cl}.  By the same mechanism, the anisotropy of the DI
here leads to the formation of a spin texture across the vortex core
that approaches a tangential disgyration as shown in
Fig.~\ref{fig:spin3}.  In a condensate where
$\xi_F$, and therefore the vortex core, is not small on the scale of
experimental resolution, this effect could be observed in experiment.

\section{Conclusions}
\label{sec:conlusions}

We have demonstrated how DI can have
several pronounced effects on the internal structure of vortices in spinor
BECs, and determined and analyzed the stable core structure of
singular vortices in a
spin-3 $^{52}$Cr condensate. In addition to exhibiting relatively
strong dipolar interactions, the $^{52}$Cr interaction parameters
predict a ground-state order parameter exhibiting a hexagonal
point-group symmetry that makes it a candidate for experimental
observation of non-Abelian vortices.

We used a spin-1 model system to analyze the underlying physical
principles and established simple criteria that determine the defect core
structure in the
presence of DI.   We have shown how a new characteristic
length scale arising from the DI adds to the hierarchy of
healing lengths to restrict the size of singular-vortex cores when DI
and $s$-wave scattering both favor the ground-state spin condition
(e.g., in the spin-1 FM phase), but can lead to core enlargement when
they compete, as in the spin-1 polar phase.
 These dipolar effects arise from generic
properties of the interaction and the spinor system and our results can
therefore be expected to apply generally across condensates of atoms
with different atomic spin.

We have also shown how the spin ordering induced by the anisotropy of the DI has
several different manifestations in both singular and nonsingular vortices.
These include a nontrivial interaction dependence of the angular momentum
carried by a coreless vortex, arising as a result of competition between dipolar
spin ordering and rotation, as well as the deformation of a singular vortex when
the DI is modified by a sufficiently strong magnetic field. The spin ordering
can also give rise to internal spin textures in the superfluid vortex cores with
nonzero spin, which are potentially observable in non-Abelian vortices in
$^{52}$Cr condensates. Similar studies of the effects of DI could be extended
beyond vortices to other more complex defects and
textures~\cite{ruostekoski_prl_2003,tiurev_pra_2016,savage_prl_2003} in which
case their symmetries and stability properties could be altered.

\begin{acknowledgments}
We acknowledge financial support from the EPSRC. The numerical results
were obtained using the
Iridis~4 high-performance computing facility at the University of
Southampton. We acknowledge discussions with T.~P.~Simula.
\end{acknowledgments}

\appendix

\section{Explicit derivation of Fourier Transforms}
\label{app:ft}

In numerical computations, it is convenient to calculate DI
contributions in Fourier space, where the convolution integrals
arising from the long-range nature of the DI become a simple matter of
multiplication of Fourier transforms.  However, the
$\delta$-function contribution to the magnetic dipole field and its
absorption into the $s$-wave interaction introduces subtleties into
the derivation of these Fourier transforms.
Here we first carefully derive the Fourier transform of the magnetic dipole
field, keeping track of all contributions, and show how it yields
Eq.~\eqref{eq:q-transf} after explicitly subtracting the
contact-interaction part. We then show how the Fourier transform is
modified by the introduction of a spherical long-range cut-off and
also indicate how the corresponding derivation is modified when the DI
is averaged over a rapid Larmor precession.

The magnetic field $\B(\rr)$ from a magnetic point dipole $\m$ at the origin is
given by~\cite{jackson}
\begin{equation}
  \label{eq:dipole-field}
  \B(\rr) = \frac{\mu_0}{4\pi r^3}[3\rhat(\m\cdot\rhat)-\m]
  + \frac{2\mu_0}{3}\m\delta(r),
\end{equation}
where $\rhat=\rr/r$.
We can write Eq.~\eqref{eq:dipole-field} on tensor form as
\begin{equation}
  \label{eq:dipole-field-tensor}
  B_\alpha = \frac{\mu_0}{4\pi} \sum_\beta\Btens_{\alpha\beta}m_\beta,
\end{equation}
where
\begin{equation}
  \label{eq:b-tensor}
  \Btens_{\alpha\beta}(\rr) =
    \frac{3\hat{r}_\alpha\hat{r}_\beta-\delta_{\alpha\beta}}{r^3}
    + \frac{8}{3}\delta_{\alpha\beta}\delta(r).
\end{equation}
The $\delta$-function contribution to the field follows from
\begin{equation}
  \label{eq:dipole-integral}
  \int_\delta\B\,d^3r = \frac{2\mu_0}{3}\m,
\end{equation}
with the convention
that the integral of the first term in Eq.~\eqref{eq:dipole-field}
vanishes on any infinitesimal sphere $\delta$ surrounding the point dipole
(integrating over angles first to make the integral converge).
In writing the dipolar Gross-Pitaevskii Hamiltonian,
Eq.~\eqref{eq:hamiltonian-density}, however, this contact-interaction
contribution is absorbed by the $s$-wave interaction, yielding
an effective field
\begin{equation}
  \label{eq:effective-field}
  \B^\prime(\rr) \equiv \B(\rr)-\frac{2\mu_0}{3}\m\delta(\rr)
\end{equation}
that corresponds to the tensor
\begin{equation}
  \Qtens_{\alpha\beta} \equiv \Btens_{\alpha\beta}
  - \frac{8}{3}\delta_{\alpha\beta}\delta(r)
  = \frac{3\hat{r}_\alpha\hat{r}_\beta-\delta_{\alpha\beta}}{r^3},
\end{equation}
appearing in Eqs.~\eqref{eq:b-vector} and \eqref{eq:q-tensor} and whose
Fourier transform is needed in Eq.~\eqref{eq:b-transf}.

In finding the Fourier transforms of $\Btens$ and $\Qtens$ we need to
ensure that the $\delta$-function contribution and its subtraction are
correctly accounted for.
In order to compute the Fourier transform, it is convenient to rewrite
the dipole field as
\begin{equation}
  \label{eq:dipole-rewritten}
  \B = \frac{\mu_0}{4\pi}(\m\times\grad)\times\grad\frac{1}{r}.
\end{equation}
In the tensor notation, this corresponds to rewriting $\Btens$ as
\begin{equation}
  \label{eq:dipole-rewritten-tensor}
  \Btens_{\alpha\beta} = \sum_{\gamma\mu\nu}
  \epsilon_{\alpha\gamma\mu}\epsilon_{\gamma\beta\nu}\partial_\nu\partial_\mu
  \frac{1}{r},
\end{equation}
where $\epsilon_{\alpha\beta\gamma}$ is the fully antisymmetric
Levi-Civita tensor.
It is straightforward to check that Eq.~\eqref{eq:dipole-rewritten}
gives the correct
field away from the origin [corresponding to the first term of
Eq.~\eqref{eq:dipole-field}].  However, we now need to check that the
condition~\eqref{eq:dipole-integral} is satisfied.
To do this, we first rewrite the
integral of $\B$ as a surface integral by rearranging the
cross products and using the divergence theorem:
\begin{equation}
  \begin{split}
  I &\equiv \int_\delta
       \frac{\mu_0}{4\pi}(\m\times\grad)\times\grad\frac{1}{r}\,d^3r\\
    &= -\int_{\partial\delta}
       \frac{\mu_0}{4\pi}\rhat\times(\m\times\grad)\frac{1}{r}\,dS.
  \end{split}
\end{equation}
Then using $\grad(1/r)=\rr/r^3$ and writing $dS = r^2 d(\cos\theta)d\phi$
in spherical coordinates, the integral becomes
\begin{equation}
  \label{eq:dipole-rewritten-integral}
  \begin{split}
    I &= \frac{\mu_0}{4\pi} \int_{\partial\delta}
         \rhat\times(\m\times\rhat)\,d(\cos\theta)d\phi\\
      &= \frac{\mu_0}{4\pi} \int_{\partial\delta}
          [-\rhat(\m\cdot\rhat)+\m]\,d(\cos\theta)d\phi,\\
      &= \frac{2\mu_0}{3}\m,
  \end{split}
\end{equation}
in agreement with
Eq.~\eqref{eq:dipole-integral}. We have thus verified that
Eqs.~\eqref{eq:dipole-rewritten} and
\eqref{eq:dipole-rewritten-tensor} correctly yield the magnetic dipole
field, Eq.~\eqref{eq:dipole-field}, including the $\delta$-function
contribution~\footnote{
  Note that while
  Eq.~\eqref{eq:dipole-integral} is here automatically fulfilled when
  the magnetic dipole field is written on the form of
  Eq.~\eqref{eq:dipole-rewritten}, the same is not true for the
  corresponding equations in the case of an electric dipole.  In that case, a
  $\delta$-function correction to the electric-field analogs of
  Eqs.~\eqref{eq:dipole-rewritten} and
  \eqref{eq:dipole-rewritten-tensor} is necessary.}

We can now proceed to find the Fourier transform of the full magnetic
dipole field as
\begin{equation}
  \label{eq:dipole-transf-1}
  \begin{split}
    \tilde{\B}(\kk) &=
    \int e^{-i\kk\cdot\rr}
    \frac{\mu_0}{4\pi}(\m\times\grad)\times\grad\frac{1}{r} \, d^3r \\
    &= -\frac{\mu_0}{k^2}(\m\times\kk)\times\kk,
  \end{split}
\end{equation}
where we have first used the vector identity
\begin{equation}
  \label{eq:integral-identity}
  \int_V f(\mathbf{v}\times\grad)\times\grad g \, d^3r
  = \int_V g(\mathbf{v}\times\grad)\times\grad f \, d^3r,
\end{equation}
and then find the remaining integral as
\begin{equation}
  \begin{split}
    \int \frac{e^{-i\kk\cdot\rr-\mu r}}{r}\,d^3r
    &= \frac{2\pi}{ik}\left(\frac{1}{-ik-\mu}-\frac{1}{ik-\mu}\right)\\
      &\stackrel{\mu\to0}{\longrightarrow} \frac{4\pi}{k^2},
  \end{split}
\end{equation}
using the convergence factor $\mu$, to yield the right-hand side of
Eq.~\eqref{eq:dipole-transf-1}.
Rewriting Eq.~\eqref{eq:dipole-transf-1} using vector identities we
arrive at
\begin{equation}
  \label{eq:dipole-tranfs-2}
  \tilde{\B}(\kk) = -\mu_0\left[\khat(\m\cdot\khat)-\m\right].
\end{equation}
From the tensor notation $\tilde{B}_\alpha(\kk) =
[\mu_0/(4\pi)]\sum_{\beta}\BtensF_{\alpha\beta}m_\beta$,
it follows immediately that
\begin{equation}
  \label{eq:btens-transf}
  \BtensF_{\alpha\beta}(\kk)
  = -4\pi(\hat{k}_\alpha\hat{k}_\beta-\delta_{\alpha\beta}).
\end{equation}
When the contact part of the
interaction is absorbed by the $s$-wave interaction, however, we need to
consider instead the Fourier transform of Eq.~\eqref{eq:effective-field},
which is immediately found from linearity as
\begin{equation}
  \label{eq:dipole-tranfs-3}
  \tilde{\B}^\prime(\kk) =
  -\frac{\mu_0}{3}\left[3\khat(\m\cdot\khat) - \m\right].
\end{equation}
Writing this in tensor notation such that
$\tilde{B}^\prime_\alpha(\kk) =
[\mu_0/(4\pi)]\sum_{\beta}\QtensF_{\alpha\beta}m_\beta$
immediately yields Eq.~\eqref{eq:q-transf}.

It is common in the literature (see, e.g.,
Refs.~\cite{kawaguchi_physrep_2012,ronen_pra_2006}) to arrive at
Eq.~\eqref{eq:q-transf}, or its special case for aligned dipoles, by
ignoring the $\delta$-function contribution in
Eq.~\eqref{eq:dipole-field}, considering only $\B^\prime$ and
expressing $\Qtens$ in terms of spherical harmonics as
\begin{widetext}
\begin{equation}
  \label{eq:q-tensor-ylm}
  \Qtens(\rr) =
    -\sqrt{\frac{6\pi}{5}}\frac{1}{r^3}
    \threemat
	{\sqrt{\frac{2}{3}}Y_{2,0}(\rhat)-Y_{2,2}(\rhat)-Y_{2,-2}(\rhat)}
	{iY_{2,2}(\rhat)-iY_{2,-2}(\rhat)}
	{Y_{2,1}(\rhat)-Y_{2,-1}(\rhat)}
	{iY_{2,2}(\rhat)-iY_{2,-2}(\rhat)}
	{\sqrt{\frac{2}{3}}Y_{2,0}(\rhat)+Y_{2,2}(\rhat)+Y_{2,-2}(\rhat)}
	{-iY_{2,1}(\rhat)-iY_{2,-1}(\rhat)}
	{Y_{2,1}(\rhat)-Y_{2,-1}(\rhat)}
	{-iY_{2,1}(\rhat)-iY_{2,-1}(\rhat)}
	{-2\sqrt{\frac{2}{3}}Y_{2,0}(\rhat)}.
\end{equation}
\end{widetext}
One then makes use of the expansion of a plane wave in terms of
spherical harmonics to find
\begin{equation}
  \label{eq:ylm-transf}
  \int e^{-i\kk\cdot\rr}Y_{l,m}(\rhat)\,d\Omega
  = 4\pi(-i)^lj_l(kr)Y_{l,m}(\khat),
\end{equation}
where $j_l$ is the spherical Bessel function of order $l$.
The radial integral
\begin{equation}
  \label{eq:bessel-int}
    \int_0^\infty \frac{j_2(kr)}{r^3}\,r^2dr
    = \int_0^\infty
    u^2\left(\frac{1}{u}\frac{d}{du}\right)^2\frac{\sin u}{u}
    \frac{du}{u}
    = \frac{1}{3},
\end{equation}
where $u=kr$, can then be combined with Eqs.~\eqref{eq:q-tensor-ylm} and
\eqref{eq:ylm-transf} to find Eq.~\eqref{eq:q-transf}.  Note, however,
that this derivation drops the contact term from the outset and the
Fourier integral is made to converge only by integrating
over angles first in Eq.~\eqref{eq:ylm-transf}.
Nevertheless, having now established that the use of
Eqs.~\eqref{eq:ylm-transf} and \eqref{eq:bessel-int} does in fact give the
correct result when the contact part of the DI is absorbed in the
$s$-wave interaction, this provides a convenient way to include the
long-range cut-off that is necessary in numerical computations, as
this affects the Fourier integral only away from the origin.  Then
truncating the DI at a radius $R$, Eq.~\eqref{eq:bessel-int} becomes
\begin{equation}
  \label{eq:bessel-int-trunc}
    \int_0^R \frac{j_2(kr)}{r^3}\,r^2dr
    = \frac{1}{3} + \frac{kR\cos(kR) - \sin(kR)}{(kR)^3},
\end{equation}
from which Eq.~\eqref{eq:q-transf-trunc} follows immediately.
This is the immediate generalization of the
spherical cut-off found by Ronen et al~\cite{ronen_pra_2006} for
aligned dipoles in a scalar BEC to free dipoles.

Finally we consider the effective DI arising when the interaction is
averaged over the Larmor precession period in the presence of a
sufficiently strong magnetic field.  We can rewrite the corresponding
tensor $\Qtens^{\mathrm{L}}$ given by Eq.~\eqref{eq:q-tensor-l} as
\begin{equation}
  \label{eq:q-tensor-l-y20}
  \Qtens_{\alpha\beta}^\mathrm{L}(\rr)
  = \frac{Y_{2,0}(\rhat)}{r^3}
    \frac{3\delta_{z\alpha}\delta_{z\beta}-\delta_{\alpha\beta}}{2},
\end{equation}
and proceed as above.  Then, from Eqs.~\eqref{eq:ylm-transf} and
\eqref{eq:bessel-int-trunc},  Eq.~\eqref{eq:q-transf-l-trunc} follows
immediately.

\section{Relative strength of DI}
\label{app:interaction-strengths}

For our studies, we have employed a suitably simple spin-1 model system to
illustrate the physics arising from DI.  The DI coupling constant $\cdd$ is then
regarded as a freely variable parameter.  For physical atoms, however, the
atomic dipole moment, and therefore $\cdd$, is a fixed quantity. In this
Appendix we briefly outline how the dipolar nonlinearity (as given in
Fig.~\ref{fig:spin-vortex-core}) can also be varied within the same spin-1 model
by adjusting trap parameters and scattering lengths.

The effective nonlinearities in the GPEs, expressed in dimensionless units as in
Fig.~\ref{fig:spin-vortex-core}, scale with the number of atoms $N$ in the
condensate and the trap frequency $\omega$ as $\sim N\omega^{1/2}$.  Therfore
the DI nonlinearity $N\cdd/(\hbar\omega\ell^3)$, used in the figure, can be
varied also for constant $\cdd$ by adjusting $N$ and/or $\omega$. We can
illustrate this principle using $^{87}$Rb as a particular example:
$Nc_0=10^4\hbar\omega\ell^3$, in Fig.~\ref{fig:spin-vortex-core} then
corresponds to $N \simeq 5\times10^5$ atoms in an $\omega \simeq 2\pi \times
10$Hz trap, and the physical magnetic dipole moment gives $N\cdd \simeq
4.2\hbar\omega\ell^3$.  By doubling the atom number to $N=10^6$ and increasing
the trap frequency to $\omega \simeq 2\pi \times 60$Hz, we can reach $N\cdd
\simeq 20\hbar\omega\ell^3$.

However, adjusting the trap parameters also scales the contact-interaction
nonlinearity $Nc_0/(\hbar\omega\ell^3)$.  This can be prevented by
simultaneously suppressing the contact-interaction coupling constant $c_0$.  The
suppression may be achieved using ac Stark shifts to access Feshbach resonances
for the $s$-wave scattering lengths without freezing out the atomic
spin~\cite{papoular_pra_2010,borgh_prl_2012}.  In the $^{87}$Rb example, the
required suppression is on the order of a factor $\sim 5$. Using these
techniques, a strongly dipolar BEC could also be achieved using $^{85}$Rb where
scattering lengths are tunable across orders of
magnitude~\cite{cornish_prl_2000}.


%

\end{document}